\documentclass[]{aa}
\usepackage{graphicx}
\usepackage{soul}
\usepackage[usenames,dvipsnames]{color}

\begin{document}
\def\mycolumnwidth{8.0cm}
\def\myhalfcolumnwidth{3.84cm}
\def\myseventhcolumnwidth{1.04cm}
\def\myttcolumnwidth{5.20cm}
\def\mytncolumnwidth{0.32cm}
\titlerunning{Detection of comet Hale-Bopp at 30.7 AU}
\title{Frozen to death? -- Detection of comet Hale-Bopp at 30.7 AU}

\authorrunning{Gy. M. Szab\'o, K. S\'arneczky,L. L. Kiss}
\author{
Gy. M. Szab\'o$^{1,2,3}$,
K. S\'arneczky$^{1}$,
L. L. Kiss$^{1,4}$
}
\institute{
Konkoly Observatory, Konkoly-Thege Mikl\'os \'ut 15-17, H1121 Budapest, Hungary
\and 
E\"otv\"os Fellow at the Astronomy Dept., University of Texas at Austin, 78712 TX, USA
\and
Dept. of Exp. Physics \& Astronomical Observatory,
      University of Szeged, 6720 Szeged, Hungary
\and
Sydney Institute for Astronomy, School of Physics, University of Sydney, NSW 2006, Australia
}


\abstract
{Comet Hale--Bopp {(C/1995 O1)} has been the single most significant comet encountered by modern astronomy, still having displayed significant activity at 25.7 AU solar distance in late 2007. It is a  {puzzling} question when and where
this activity will finally cease.} 
{Here we present new observations with the ESO 2.2m telescope at La Silla to check the activity of Hale--Bopp at 30.7 AU solar distance.} {On 2010-12-04, 26 CCD images were taken with 180 s exposure times for photometry and morphology.} 
{The comet was detected in $R$ and had a total brightness of 23\fm3$\pm$0\fm2, {referring
to an absolute brightness of $R$(1,1,0)=8\fm3}.
The profile of the coma was star-like {at a seeing of 1\farcs9}, without any evidence of a coma or tail extending farther than 2\farcs5 (=55,000 km  {in projection}) and exceeding 26.5 mag/$arcs^2$ surface brightness.} {The measured total brightness corresponds to a relative total reflecting surface, $a_RC$, of 485 km$^2$, nine times less than three years before. The calculated $a_RC$ value would imply a nucleus with 60--65 km radius assuming 4\%{} albedo. This size estimate is in significant contradiction with the previous results scattering around 35 km. Therefore we suggest that the comet may still be in a low-level activity, despite the lack of a prominent coma. Alternatively, if the nucleus is already dormant, the albedo should be as high as 13\%{}, assuming a radius of 35 km.
With this observation, Hale--Bopp has been the most distant comet ever observed, far beyond the orbit of Neptune.}

\date{}

\maketitle

\section{Introduction.}
According to the paradigm of cometary activity, matter production decreases fast after the {rapid decrease} of water sublimation at 3 AU (Fern\'andez 2005). A few exceptional comets, however, displayed activity far beyond 3 AU (e.g. Lowry et al. 1999; Lowry \&{} Fitzsimmons 2001, 2005; Lowry \&{} Wiessman 2003; Snodgrass et al. 2006, 2008; Mazzotta Epifani et al. 2006, 2007). There are known examples of active long period comets (Szab\'o et al. 2001a, 2002, Meech et al. 2004, Mazzotta Epifani et al. 2009, Ivanova et al. 2011) and Centaur objects (e.g. Meech et al. 1997, Rousselot 2008, Jewitt 2009). 
The exact location of the snow line is very important in understanding the water reservoirs in the Solar System, and therefore the structure and evolution of extrasolar systems, too.

In Szab\'o et al. (2008) we have shown that C/1995 O1 (Hale--Bopp) was the most distant active comet ever observed, with a prominent coma at 25.7 AU solar distance. This surprising observation raised the question where this activity will cease. Observing Hale--Bopp in a completely frozen state would be also extremely important because a thick coma was constantly present during the entire apparition. The coma obscured the nucleus which was not observed directly. Lacking photometric data of the bared nucleus, its size, one of the most important input parameter in activity models remained uncertain.

Here we present new observations of Hale--Bopp at 30.7 AU solar distance. In late 2010, the comet exhibited evidence for the cessation of the matter production. In this paper we present a photometrical and morphological analysis of these observations.

\section{Observations}

We made new $R_C$ filtered observations with the 2.2 m ESO telescope at the La Silla site in Chile, on December 4, 2010, {with sidereal tracking}. We took 26$\times$180~s exposures with an image scale of 0\farcs24/pixel. The seeing was $\approx$1\farcs9, mostly due to the airmass of 1.9--2.1, which was caused by the comet's unfavourable celestial position close to the Southern pole.

\begin{figure}
\begin{center}
\includegraphics[width=\columnwidth]{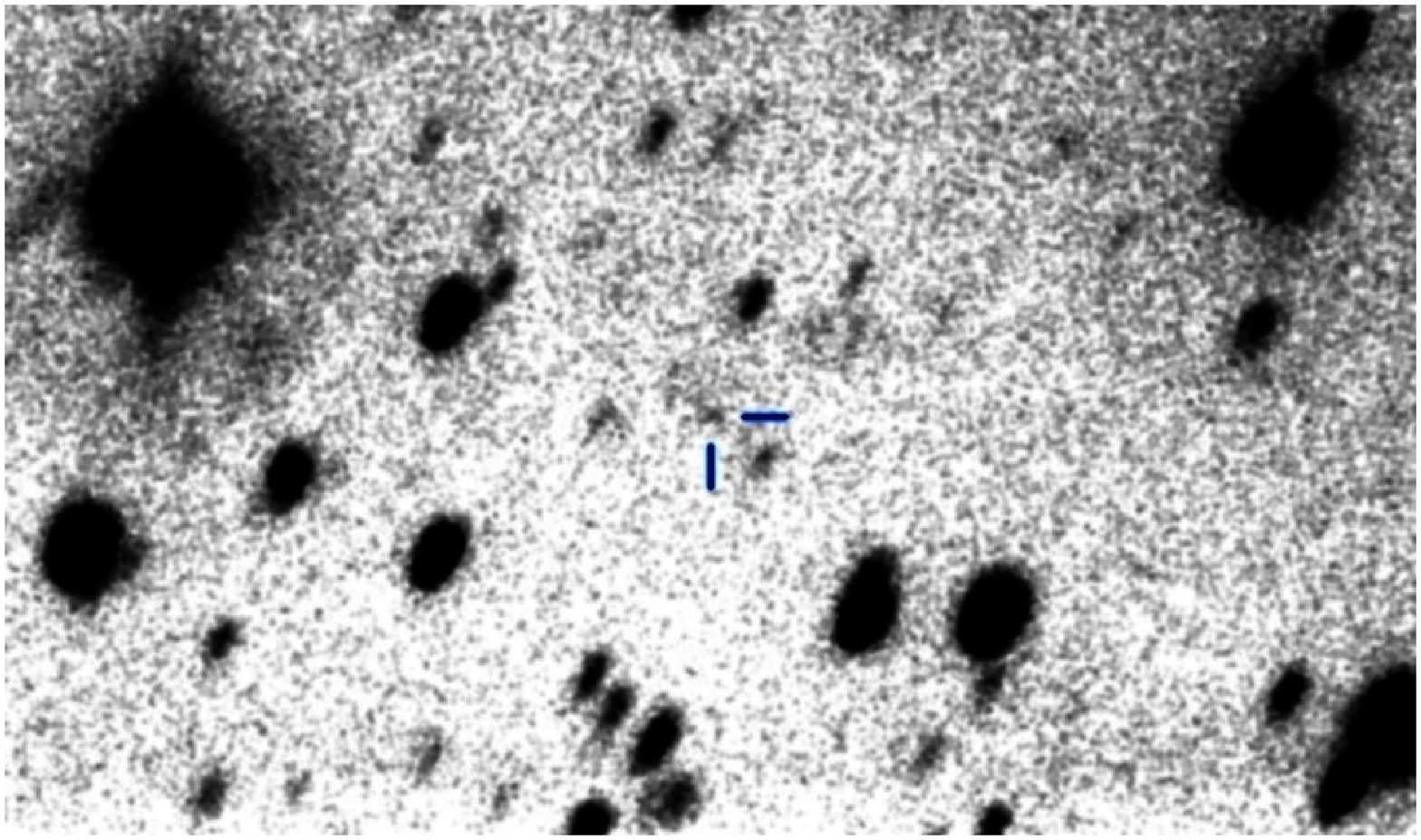}
\medbreak
\includegraphics[width=\columnwidth]{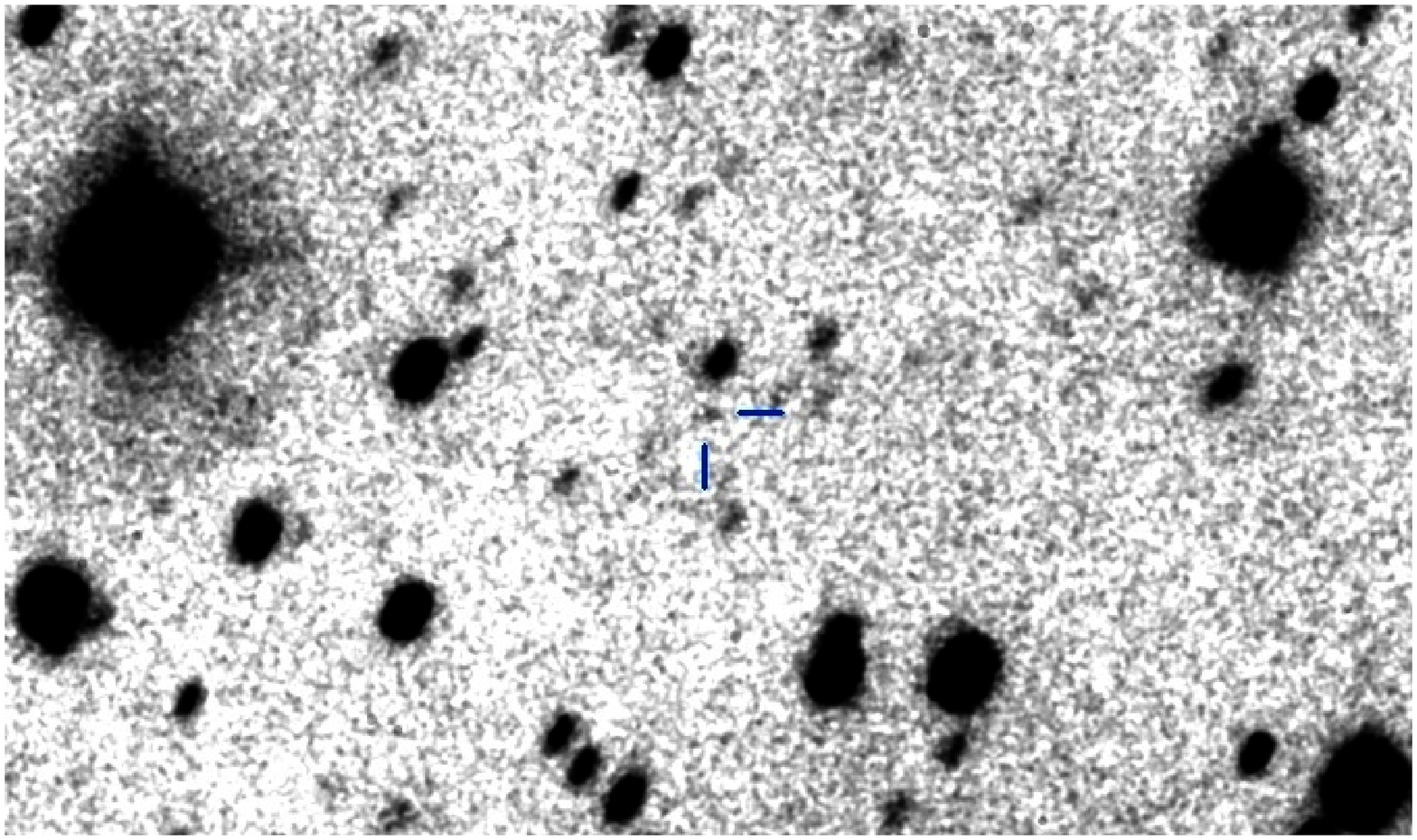}
\end{center}
\medbreak
\includegraphics[width=\columnwidth]{hbpath.epsi}
\caption{Upper images: comet Hale--Bopp at 30.7 AU solar distance, on December 4, 2010. The field of view is 2\farcm67$\times$1\farcm57; north is up, east is to the left. The brightest star in the image is 2MASS J00395120-8632270, which has an R magnitude of 12.93$\pm0.06$. Lowest panel: Comparison of the measured astrometric positions (open dots with errors) and the JPL predictions (solid dots). Error bars denote the 2-sigma level confidence intervals of the 
measured positions.}
\label{zsaner}
\end{figure}

The images were corrected in a standard fashion, including bias and flat field correction. We aligned and co-added the images by fitting a coordinate grid to the stars, yielding a ``star field'' image for photometric calibrations. The images were then re-aligned with respect to the proper motion of the comet, to get untrailed ``comet'' images. In this step, the MPC ephemerides at the time of each observation were used to match the individual frames. Fig. \ref{zsaner} shows the ``comet'' images on December 4, 2010, after co-adding the images taken  {between 5:41--6:48 UT (top panel), and between 7:08--8:10 UT (lower panel)}. The proper motion of the comet confirms its identification.

Hale--Bopp was detected {very} slightly northwest from the positions predicted for the La Silla site at the time of observations (Table 1, astrometry part; Fig. 1 lowest panel). The mean deviations and their $rms$ scatter in right ascension and declination are 0\farcs59$\pm$0\farcs06 and 0\farcs25$\pm$0\farcs20, respectively. The confidence interval of the measured positions was tested by astrometry of artificial sources: with {\sc addstar} routine in IRAF, we added 18 artificial comets to the image with the same PSF and brightness as the comet image itself. Because of the noise and sampling, the astrometry of these artificial comets resulted in slightly different positions than where they were inserted. The standard error of these differences resulted to be 0\farcs23 in both coordinates, which can be considered as the 1-sigma confidence level error of the measured positions of Hale-Bopp. {NASA JPL Horizon reports 3-$\sigma$ uncertainties on the position of 0\fs 45 and 0\farcs10 in RA and DEC, respectively.} The measured {very marginal} differences may be attributed to non-gravitational forces from the distant activity of the comet. 

For the photometry, we used local comparison stars taken from the USNO-B catalog. To check the consistency of the magnitude zero-point, another method was also used, based on synthetic $R_C$ magnitudes from 2MASS photometry. For calculating the synthetic $R_C$ magnitudes, we collected 2MASS photometry of all stars in the Landolt (1992) catalog and followed the method of least-square fitting of the color indices (see Bilir et al. 2008, who give transformations from 2MASS magnitudes to Sloan $g−r$ and $r−i$ colours; we followed the same procedure to determine a transformation from 2MASS magnitudes to $R_C$ magnitude). We excluded all stars with $K>13$ from the fitting, and also restricted $J-K<0.4$  {where the color-color distributions were narrow and linear}. We then determined
\begin{equation}
R_C = J - 0.766(J-H) + 2.364  (J-K) + 0.213 \pm 0.055,
\end{equation}
valid for stars $(J-K < 0.4)$. We identified 15 appropriate stars in our images which had 2MASS photometry. Eq 1 has been applied to these stars, and finally the photometric zero-point of the average of 180 s images resulted to be 24\fm42$\pm$0\fm06 magnitudes. This result is absolutely consistent with a zero-point of 24\fm45$\pm$0.05, derived from the $R$ photometry presented in the USNO-B catalog. The 4\%{} difference of the two different zero-points are much less than the photon noise of the comet image itself. We finally adopted a zero-point of $24.43$ magnitudes {at the average airmass of $\approx$2 where the observations were taken. This ZP is consitent with the standard value of WFI of $\approx$24.4--24.6 in the Zenith (chip \#{}6), and the average $R$ extinction coefficient of 0.07--0.12.} 

{The comet and the comparison stars were measured in an aperture which had 2\farcs9 diameter, i.e. 1.5-times the FWHM of stellar profiles. This is quite close to the aperture of the optimal S/N (1.3$\times$FWHM, Szab\'o et al. 2001b), and contains 85\%{} of the starlight. The total photon noise (star+background) was 0\fm08 (1-$\sigma$ confidence interval).  Because the comparison stars were measured with the same aperture, there was no need to aperture correction. Estimating conservatively an error of $\approx$0\fm1 in both ZP and in the systematics from aperture effects, the total error results to be 0\fm2.} Surface photometry has been applied to the each pixels of the raw ``comet'' image, i.e. without deconvolution to the seeing. Errors were estimated based on the background scatter of 1.27 ADU/pixel, or 0.07 ADU/$arcs^2$. The surface brightness profile is plotted in Fig. 2.

\section{Results}

\begin{table}
\caption{Ephemerides and astronetry of Hale-Bopp on 2010-12-04.}
{\bf Ephemerides}\\
\begin{tabular}{llllll}
\tiny $\lambda$ $[^\circ]$ &\tiny $\beta$ $[^\circ]$ &\tiny R [AU] &\tiny $\Delta$ [AU]&\tiny E $[^\circ]$ &\tiny $\alpha$ $[^\circ]$ \\
\hline
\tiny 278.8 &\tiny $-$66.9& \tiny30.7 &\tiny 31.0  &\tiny 69.5 &\tiny  1.8\\
\hline
\end{tabular}
\\
\\
{\bf Astrometry}\\
\begin{tabular}{lllll}
UT~~~~ & \multicolumn{2}{c}{RA (00$^{\rm h}$38$^{\rm m}$)~~~~~~} & \multicolumn{2}{c}{Dec (-86$^\circ$32$^\prime$)} \\
 & Pred. & Meas. & Pred. & Meas. \\
\hline
6:15& 42\fs 94& 42\fs 28 & 53\farcs9 & 53\farcs8\\
6:57 & 41\fs 17 & 40\fs 60 & 50\farcs9 & 50\farcs5\\   
7:39 & 39\fs 42 & 38\fs 73 &47\farcs8 & 47\farcs0\\
\end{tabular}
\end{table}

Compared to our previous observations, the general appearance of the comet has drastically changed during the past 3 years.
The measured total magnitude of Hale--Bopp was $R=$23\fm3$\pm$0\fm2 in 2010, {referring
to an absolute brightness of m$_R$(1,1,0)=8\fm3}. This brightness can be converted to $a_RC$, the total reflecting surface in the aperture (Eddington 1910),
\begin{equation}
a_RC = {2.22\times 10^{22} \pi R^2 \Delta^2 10^{0.4(m_{\sun} - m_{\rm comet})} \over 10^{-0.4\alpha\beta}}.
\end{equation} 
Here $m_{\sun}=-27\fm11$, the apparent $R_C$ brightness of the Sun, and the $\beta$ phase coefficient is usually assumed to be 0.04. Substituting the measured total brightness yields $a_RC\approx 485$ km$^2$. 
This value is only 11\%{} that of measured in late 2007, when Hale--Bopp was at 25.7 AU solar distance.

Unlike most comets with distant activity, the inner coma of Hale--Bopp followed a distinct, flat profile (Szab\'o et al. 2008). In 2010 we did not detect such a coma, and the profile of Hale--Bopp was not broader than that of the stars (Fig. 2). This indicated the lack of a bright, extended inner coma. The measured intensity peak of the comet had an apparent surface brightness of 24.3 mag/arcs$^2$. For comparison, the peak intensity was 20.6 mag/arcs$^2$ 3 years before, and that time a flat intensity plateau extended to $\approx$ 2\farcs5 from the nucleus. By late 2010, the bright flat coma vanished, suggesting that the matter production in 2010 was at least an order of magnitude less than in 2007. This estimate is also compatible with the measured decrement of $a_RC$ by an order of magnitude. In 2010, the comet also lacked coma or tail extending farther than 2\farcs5 and exceeding 26.5 mag/arcs$^2$ surface brightness (which is the limit of a 2-$\sigma$ detection of extended sources above the background noise). Since coma-like feature is not present in the images, we conclude that Hale-Bopp had a very faint outer coma, if any, at 30,7 AU solar distance.

Because $a_RC$ of 485 km$^2$ is in a contradiction with the expectations based on previous size estimates, we cannot exclude that a loose coma component was still present around Hale--Bopp at the time of observation. The value of $a_RC$ would refer to a total cross section of 12,000 km$^2$ of the reflecting matter, assuming 4\% albedo. A bared nucleus could only support this value if it would be as large as 62 km (radius), which is incompatible to the previous size estimates (e.g. 35 km, Weaver and Lamy 1997; 30 km, Fern\'andez 2003; 37 km, Lamy et al. 2004). Therefore, some part of the reflecting matter must be present in the form of a loose coma. This {may be} an indirect evidence for a low-level activity of Hale--Bopp, beyond the orbit of Neptune. {Another possibility is that the comet is already dormant, but
there might still be dust grains around the nucleus. If there is plenty dust in the aperture, this can significantly contribute to the total light.}

\begin{figure}
\begin{center}
\includegraphics[width=\columnwidth]{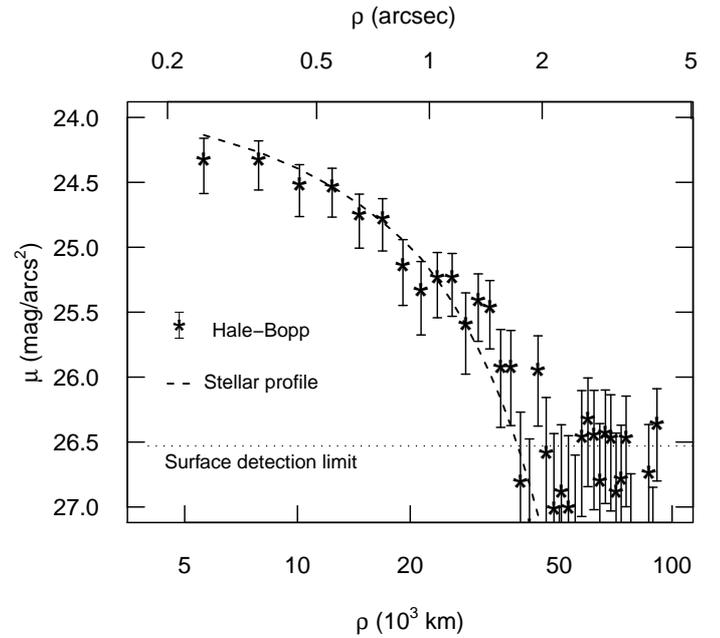}
\end{center}
\caption{Surface brightness profile of Hale--Bopp (stars), compared to the average profile of faint stars (dashed line).
The dotted line shows the 2-$\sigma$ limit of surface brightness detection in the comet image.}
\label{lcs}
\end{figure}

\begin{figure}
\begin{center}
\includegraphics[bb=69 195 455 505,width=\columnwidth]{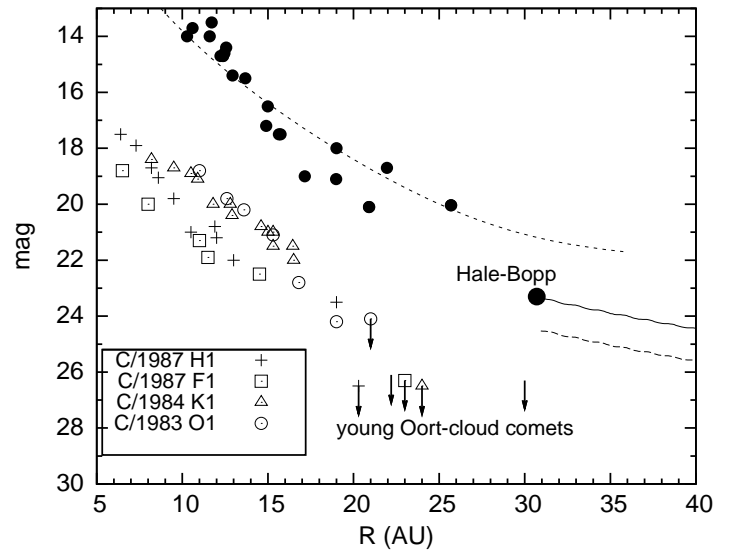}
\end{center}
\caption{The light curve of Hale--Bopp compared to observations of 6 dynamically young Oort-cloud comets (Meech et al. 2004; the last five point are upper limits). The top dashed curve shows the prediction of distant activity by Capria et al. (2002). The middle solid curve is the brightness prediction if the nucleus is {already} dormant. The lower dashed curve plots the expected brightness of a 37 km size nucleus (radius) with 4\%{} albedo (Lamy et al. 2004).}
\label{lcs}
\end{figure}

In Fig. \ref{lcs}, we plot the observed brightness of Hale--Bopp (collected from ICQ and MPC bulletins) against the solar distance. For comparison, data for 6 dynamically young Oort-comets are also plotted (Meech et al. 2004). We also show a theoretical light curve predicted from the CO production curve by Capria et al. (2002), assuming water ice crystallization heating the nucleus. In the case of previous observations, Hale--Bopp was consistently 3--5 magnitudes brighter than these Oort-comets at distances $<$15 AU. By 2010, the total brightness of the comet has diminished by 1.5--2 magnitudes. The observed sudden decrease of the total brightness, together with the disappearance of the prominent coma shows that Hale--Bopp is now on the way of freezing out: the cessation of matter production.

\section{Discussion and summary}

Our observation is worth comparing to the observations of 1P/Halley at large heliocentric distances (Hainaut et al. 2004). Observed at 28.1 AU, comet Halley had been the most distant comet ever observed until our present observations of Hale--Bopp. With the VLT an integration time of almost 9 hours has been applied, leading to the detection of Halley's nucleus with S/N=8. That time, comet Halley was a 28.1 magnitude atmosphereless body. The significant difference between these comets (over the simple fact that Hale--Bopp is larger and brighter) is how far out comet Hale--Bopp could maintain an observable matter production. The activity of comet Halley ceased at around 11 AU solar distance. Interestingly, an outburst was observed in 1991 when comet Halley suddely brightened 5 magnitudes, when the comet was at around 14.2 AU solar distance {(West et al. 1991).}

In Fig. 3, two curves illustrate the expected brightness of Hale--Bopp beyond 30 AU. The brighter state (solid line) is the predicted light curve if the nucleus is already inactive. (This is a light curve of a body with an identical absolute brightness as Hale--Bopp in late 2010.) The fainter scenario is the model of Lamy et al. (2004), plotting the brightness of a body with 37 km radius and 4\%{} albedo. Hale--Bopp was evidently brighter than this latter curve, which we invoked as an indirect evidence of a loose coma. However, there is an alternative explanation of the measured brightness. A bared nucleus with 35 km radius can exhibit the observed brightness if it has a higher albedo, approximately 13\%{}. This is significantly higher than the albedo of cometary nuclei in general and that of derived for Hale--Bopp especially (4\%{}, Lamy et al. 2004). But considering a possible re-condensation of the gases to the surface during the distant activity, it would be even plausible that the surface has been covered with a thin ice film that could increase the albedo. In the case of this scenario, the light curve of Hale--Bopp will follow the path of our brighter prediction. If the measured excess light is due to low-level activity, the light curve will converge to the fainter prediction in the future.

In Szab\'o et al. (2008) we estimated the basic parameters of the activity of Hale--Bopp at 25,7 AU solar distance. We estimated its temperature was 53.1 K. This was slightly less than 54.8 K for a blackbody, due to the sublimation of $2\times 10^{19}$ CO molecules/m$^2$/s. In 30.7 AU solar distance, the equivalent temperature of a blackbody is 50.1 K. Since Hale--Bopp seems to turn into inactive state, we infer that the temperature of activity cessation is somewhere between 50--53 K for a Hale--Bopp type comet.

The main results of this paper can be summarized as follows:
\begin{enumerate}
\item{} We detected comet Hale--Bopp at 30.7 AU, which is the most distant detection of a comet so far. 
\item{} {The absolute brightness of Hale--Bopp was $R$(1,1,0)=8\fm3}. The total relative reflecting surface of the 40,000 km environment around the nucleus, $a_RC$ is 485 km$^2$. This is a factor of 9 less than that we observed in 2007, and suggests the comet is near to ceasing its activity. However, this cross section is somewhat larger than one would expect for a nucleus with 30--40 km radius (unless its albedo is very high), so we suggest some light is scattered by a faint coma which is still present.
\item{} The comet has a star-like appearance, therefore we concluded the comet lacks the presence of the
thick coma which was characteristic to this comet at all observed solar distances. This situation
will finally enable a reliable coma-nucleus separation and the photometry of the nucleus with large telescopes such as VLT.
\end{enumerate}

\section*{Acknowledgments}
This project has been supported by the Hungarian OTKA Grants K76816, K83790 and MB08C 81013, the ``Lend\"ulet'' Program of the Hungarian Academy of Sciences, and the E\"otv\"os Fellowship of the Hungarian State.

\end{document}